\documentclass[aps,prb,twocolumn,showpacs,amsmath,amssymb,floatfix]{revtex4}

\usepackage{graphicx}
\usepackage{dcolumn}
\usepackage{bm}
\usepackage{epstopdf}
\begin{document}

\title{Analytical Expression for the RKKY Interaction in Doped Graphene}

\author{M. Sherafati}
\author{S. Satpathy}%

\affiliation{
Department of Physics $\&$ Astronomy, University of Missouri,
Columbia, MO 65211}

\date{\today}

\begin{abstract}

We obtain an analytical expression for the Ruderman-Kittel-Kasuya-Yosida (RKKY) interaction $J$ in electron or hole doped graphene for linear Dirac bands. The results agree very well with the numerical calculations for the full tight-binding band structure in the regime where the linear band structure is valid.
The analytical result, expressed in terms of the Meijer G-function, consists of a product of two oscillatory terms, one coming from the interference between the two Dirac cones and the second coming from the finite size of the Fermi surface.
For large distances, the Meijer G-function behaves as a sinusoidal term, leading to the result  $J \sim R^{-2} k_F \sin (2 k_F R) \{ 1 + \cos \ [(K-K').R] \}$ for moments located on the same sublattice. The $R^{-2}$ dependence, which is the same for the standard two-dimensional electron gas, is universal irrespective of the sublattice location and the distance direction of the two moments except when $k_F =0$ (undoped case), where it reverts to the $R^{-3}$ dependence. These results correct several inconsistencies found in the literature.


\end{abstract}
\maketitle

%


The Ruderman-Kittel-Kasuya-Yosida (RKKY) interaction,\cite{RKKY} which measures the coupling between two magnetic moments mediated by a background of electrons, is an important characteristic of the electron system.
It has been extensively studied for the electron gas in one, two, or three dimensions.
Even though graphene is a two-dimensional (2D) system, there are two important differences from the standard 2D electron gas, viz., the linear band structure and the existence of two Dirac cones in the Brillouin zone. This leads to the two characteristic momenta, the Fermi momentum $k_F$ and the momentum difference of the two Dirac points $K - K'$, both of which produce oscillatory factors of their own, leading to unusual features not found in the standard 2D electron gas, e.g, the beating of the RKKY interaction that can be controlled by a gate voltage.

Although the RKKY interaction in graphene has already been studied  by many authors,
\cite{Saremi,RKKYGraphene,FriedelGraphene} the results differ from one another,
even in the long-distance behavior, and the interference term from the Dirac cones is often missing in the results.
In this paper, we derive analytical expressions for the RKKY interaction for the linear band model, extending our earlier work for the undoped case.\cite{SashiM} The results are compared to  the same for  the tight-binding model which we also calculate from the numerical evaluation of the lattice Green's function. The linear-band and the tight-binding results agree quite well in the cases where $k_F$ lies in the linear regime ($|E_F|  \lesssim t/3$).
We find that the analytical results in the linear band approximation may be expressed as a product of the $J$ for the undoped case, which is obviously independent of $k_F$, and a new factor that depends on $k_F$ and goes to one in the limit of $k_F \rightarrow 0$, so that the results for the undoped case are correctly reproduced. The analytical results, summarized in Table I, are expressed  in terms of the Meijer G-function, whose long distance behavior is sinusoidal.

{\it Model and the method} --
We consider the nearest-neighbor  tight-binding Hamiltonian for the $\pi$-electrons in graphene including the contact interaction with two localized magnetic moments
\begin{equation}
{\cal H}  ={\cal H}_0+{\cal H}_{\text{int}},
\label{hamil}
\end{equation}
where
$
{\cal H}_0=-t \sum_{\langle ij \rangle \sigma} c^{\dagger}_{i\sigma}c_{j\sigma} + H. c.
$
is the tight-binding Hamiltonian, $\langle ij \rangle$ denotes summation over distinct pairs of nearest neighbors, $t \approx 2.56$ eV,\cite{Nanda-Bilayer}      $\sigma$ is the spin index,
and the interaction term between the localized spins $\vec S_p$ and the itinerant electron spins $\vec s_p$ is given by
$
{\cal H}_{\text{int}} = - \lambda (\vec S_1 \cdot \vec s_1+\vec S_2 \cdot \vec s_2).
$
In the linear response theory, the interaction energy
may be written in the Heisenberg form
\begin{equation}
E(R) = J_{\alpha \beta} (R) \vec S_1 \cdot \vec S_2,
\label{RKKYE}
\end{equation}
where the sublattice indices  and the positions of the two moments are $(\alpha , 0)$ and   $(\beta, R)$,
$
J_{\alpha \beta} (R) = 4^{-1} \lambda^2 \hbar^2  \chi_{\alpha \beta} (0,R)
$
is the RKKY interaction and $ \chi_{\alpha \beta} (r,r^\prime) \equiv  \delta n_\alpha(r) / \delta V_\beta(r')$ is the susceptibility.
Note that  $R$ denotes the position of the atom and $\it not$ the position of the cell in which it is located; they differ by the basis vector of the atom in the unit cell.

The susceptibility can in turn  be computed from the unperturbed  Green's function
\cite{Horiguchi,SashiM,Nanda-Vacancy,Sherafati-PSS}
\begin{equation}
\chi_{\alpha \beta} (0,R) =
- \frac{2}{\pi} \int^{E_F}_{-\infty} dE \
{\rm Im} [G^0_{\alpha \beta} (0, R, E)   G^0_{\beta \alpha} (R, 0, E)].
\label{chi}
\end{equation}
Below we evaluate this integral analytically for the linear bands and numerically for the tight-binding bands by direct integration
\begin{equation}
G^0_{\alpha \beta} (R,0, E) = \frac{1}{\Omega_{BZ}}\int d^2k e^{i k \cdot R}   G^0_{\alpha \beta} (k, E),
\label{GF2}
\end{equation}
of the momentum-space Green's function
\begin{equation}
G_{\alpha \beta} ^0 (k,E)
=  \frac{E+i \eta + {\cal H}_k}{(E+i \eta )^2 - |f(k)|^2}.
\end{equation}
Here
$
{\cal H}_k=
 \left( \begin{array}{cc}
0 &  f(k)\\
f^*(k) & 0
\end{array}\right)
$
is the graphene tight-binding
Hamiltonian   in the momentum space and the Bloch sum
$f(k) = -t \ (e^{i  k \cdot  d_1} + e^{i  k \cdot  d_2} + e^{i  k \cdot  d_3} ) $, where $d_1$, $d_2$ and $d_3$

\begin{widetext}

\begin{table} [t]
\caption{Summary of the RKKY interaction in graphene for both the doped  ($k_F \ne 0$) and the undoped case  ($k_F = 0$), given as a product of the terms: $J_{\alpha \beta} = \alpha_C \alpha_D \alpha_F$. The long-distance behavior is obtained by replacing $\alpha_F$ with $\alpha'_F$. Here $C \equiv 9 \lambda^2 \hbar^2/ (256 \pi t)$, $C' \equiv -\lambda'^2 V^2 m^*/8 (N \pi \hbar)^2$, $x_D=(K-K^\prime) \cdot R$, $x_F=k_F R$,  $\theta_R$ is the angle of the position vector $R$ made with $K'-K$ direction, where $K'$ and $K$ are two adjacent Dirac points in the Brillouin zone. The results for the standard two-dimensional electron gas (2DEG) \cite{RKKY-2D} are also shown,
where we have rederived the long-distance behavior.
}
\begin{center}
\begin{tabular}{c|ccccc}
\hline

Sublattices&$k_F$&$\alpha_C$&$\alpha_D$&$\alpha_F$&$\alpha'_F$\\
$\alpha, \beta $ &    & Prefactor&Dirac-cone factor  & Fermi factor & Long-distance bevavior of $\alpha_F$\\
\hline
\hline
$AA$&0&$-C(R/a)^{-3}$&$1 + \cos x_D$&$1$&$1$\\
$AA$&$k_F$&$-C(R/a)^{-3}$&$1 + \cos  x_D $&$1+8\pi^{-1/2}x_F M(x_F)$&$\pi^{-1}[2 \cos(2x_F)+8x_F \sin(2x_F)]$\\
$AB$&0&$3C(R/a)^{-3}$&$1 + \cos (x_D+\pi-2 \theta_R)$&$1$&$1$ \\
$AB$&$k_F$&$3C(R/a)^{-3}$&$1 + \cos (x_D+\pi-2 \theta_R)$&$1-8 (9 \pi)^{-1/2}x_F M'(x_F)$&$(3\pi)^{-1}[10 \cos(2x_F)+8x_F \sin(2x_F)]$ \\
\hline
2DEG&$k_F$&$C'R^{-2}$&1&$x_F^2[J_0(x_F)Y_0(x_F)+J_1(x_F)Y_1(x_F)]$&$(4 \pi x_F)^{-1}[\cos(2x_F)-4x_F \sin(2x_F)]$ \\
\hline
\end{tabular}
\end{center}
\end{table}
\end{widetext}
%
%
are the three nearest-neighbor position vectors. Note
from expressions following Eq. (\ref{RKKYE}) that the Friedel oscillations\cite{FriedelGraphene} $\delta n_\alpha (r)$ in the charge density  induced  by a $\delta$-function potential is proportional to $J_{\alpha\beta}$ as well.

{\it Moments on the same sublattice} -- Using methods discussed
in our previous work \cite{SashiM}, the Green's functions as well as the susceptibility  can be evaluated both for the linear-band approximation and for the full tight-binding bands. For the linear-band case and for moments on the same sublattice, the result is
\begin{equation}
\chi_{AA} (0,R) = I_{AA} (R) \times \{1 + \cos [(K-K^\prime) \cdot R ] \},
\label{chi-AA}
\end{equation}
where
\begin{equation}
I_{AA} (R) = - \frac{4}{\pi} \int_{-\infty}^{E_F}  dE \  {\rm Im}  \ [g_{AA} (R, E)]^2 ,
\label{IAA}
\end{equation}
$g_{AA}(R,E) =  -2 \pi E v_F^{-2} \Omega_{BZ}^{-1} K_0(-i E R/v_F), $
$K_0$ is the modified Bessel function of the second kind, $v_F=3 t a/2$ is the Fermi velocity, $a$ is the bond length and $\Omega_{BZ}$ is the area of the Brillouin Zone.
Now we split the integral in Eq. (\ref{IAA}) into two parts, viz., $ \int_{-\infty}^{E_F}= \int_{-\infty}^{0}+ \int_{0}^{E_F}$, where the first term accounts for the valance electrons (undoped case) and the second for the conduction electrons, so that
\begin{equation}
I_{AA} (R) = \frac{8 \pi^3}{\Omega_{BZ}^2 v_F} R^{-3} [I_0 +
                  \int_0^{k_FR} dz \ z^2 J_0 (z) Y_0 (z)],
\label{IAA-tot}
\end{equation}
where
$ I_0 = -\int_0^\infty dy \ y^2 J_0 (y) Y_0 (y) = -1/ 16$,\cite{SashiM}
$y=-ER/v_F$ for the valance band ($E<0$), $z=ER/v_F$ for the conduction band ($E>0$) and $J_0$ and $Y_0$ are the Bessel and Neumann functions with real arguments and $k_F$ is the Fermi momentum.

\begin{figure} [tp]
\centering
\includegraphics[width=6.0cm]{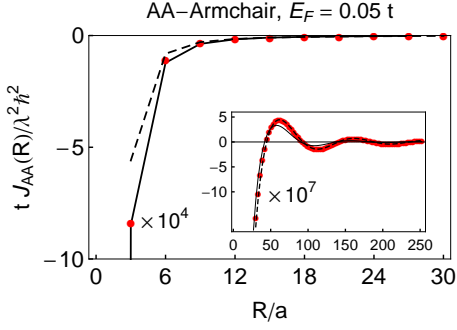}
\caption{(Color online) RKKY interaction $J_{AA}$ obtained for the tight-binding bands (solid line), compared to the
RKKY expression  involving the Meijer G-function Eq. (\ref{JAAdoped}) (red dots)  and its long-distance limit Eq. (\ref{JAAlrg}) (dashed line).
}
\label{MGvsNum2}
\end{figure}

The remaining integral in Eq. (\ref{IAA-tot}) may be expressed in terms of the Meijer G-functions. The product of the Bessel and the Neumann functions can be written as
\begin{equation}
z^\mu J_\nu (z) Y_\nu (z)=-\pi^{-1/2} G_{1,3}^{\,2,0} \!\left( \left. \begin{matrix} \frac{\mu+1}{2} \\ \frac{\mu}{2},\frac{\mu}{2}+\nu, \frac{\mu}{2}-\nu \end{matrix} \; \right| \, z^2 \right),
\label{MeijerG-Bessel}
\end{equation}
and using the integral tables \cite{Meijer} along with $\mu=2$, $\nu=0$, and the new variable $x=z^2(k_FR)^{-2}$,  the result is
\begin{eqnarray}
\begin{split}
& \int_0^{k_F R} dz \ z^2 J_0 (z) Y_0 (z)=  \frac{-\pi^{-1/2} k_F R}{2} \int_0^{1} dx \ x^{-1/2} \times \\
& G_{1,3}^{\,2,0} \!\left( \left. \begin{matrix} \frac{3}{2} \\ 1,1, 1 \end{matrix} \; \right| \, k_F^2 R^2 x \right)
 = -\frac{k_F R}{2\sqrt{\pi}} M(k_FR),
\label{IAA-doped}
\end{split}
\end{eqnarray}
where $M(k_FR)=G_{2,4}^{\,2,1} \!\left( \left. \begin{matrix} \frac{1}{2},\frac{3}{2} \\ 1,1,1,\frac{-1}{2} \end{matrix} \; \right| \, k_F^2 R^2  \right)$ is a short-hand notation for the Meijer G-function.
Using Eqs. (\ref{chi-AA}), (\ref{IAA-tot}), and (\ref{IAA-doped}),   we arrive at our desired result, valid for the moments on the same sublattice and for the linear bands, viz.,
\begin{equation}
J_{AA}(R)=J^0_{AA}(R) \ [1+\frac{8k_F R}{\sqrt{\pi}}M(k_FR)],
\label{JAAdoped}
\end{equation}
where
$
J^0_{AA}(R) = -C \times   (R/a)^{-3}  \{1 + \cos [(K-K^\prime) \cdot R] \}
$
is the undoped exchange interaction with $C \equiv 9 \lambda^2 \hbar^2/ (256 \pi t)$.
The only approximation used here was to extend the linearity of the Dirac bands to infinity (infinite momentum cutoff); however, this approximation is in good agreement with the numerical full-band tight-binding calculations, both for the undoped case \cite{SashiM} and for the doped case if $k_F$ is small [Fig. (\ref{MGvsNum2})].

\begin{figure}
\centering
\includegraphics[width=6.5cm]{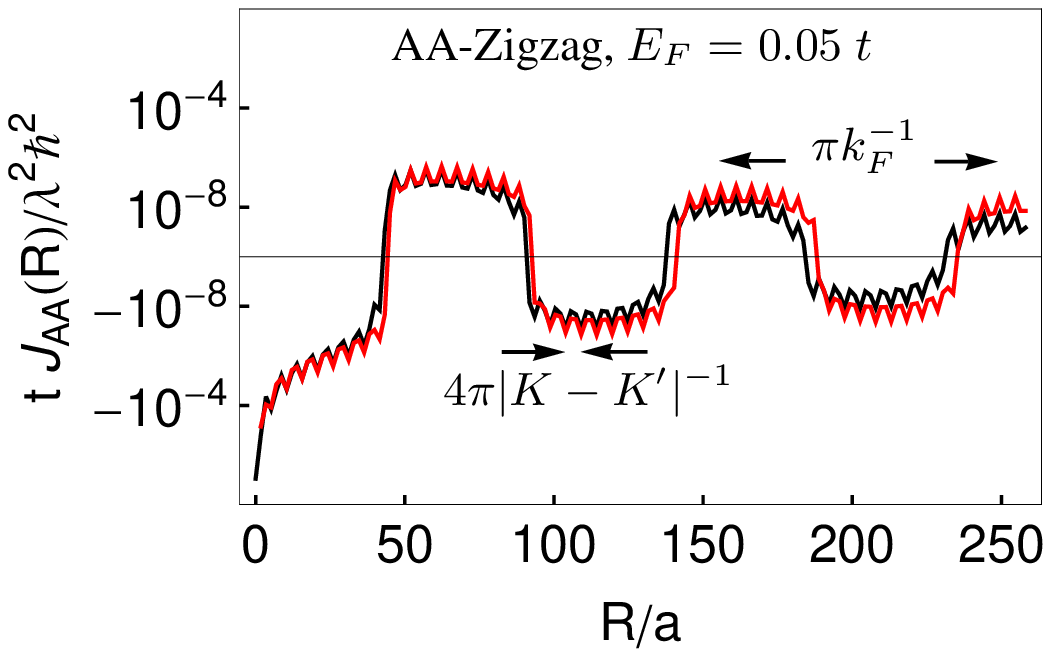}
\includegraphics[width=6.5cm]{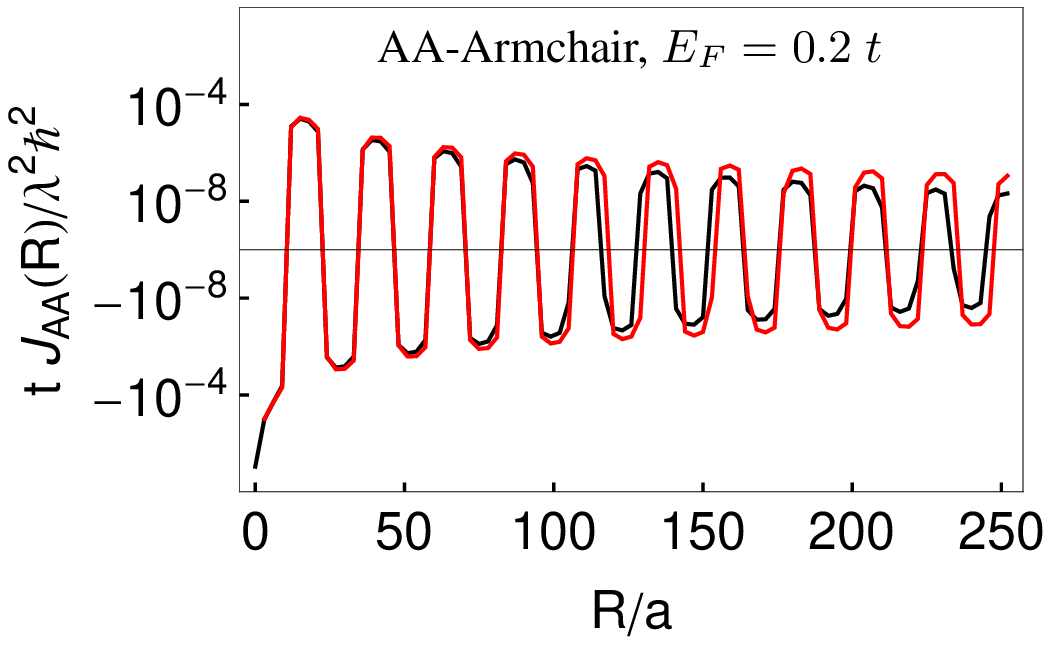}
\includegraphics[width=6.5cm]{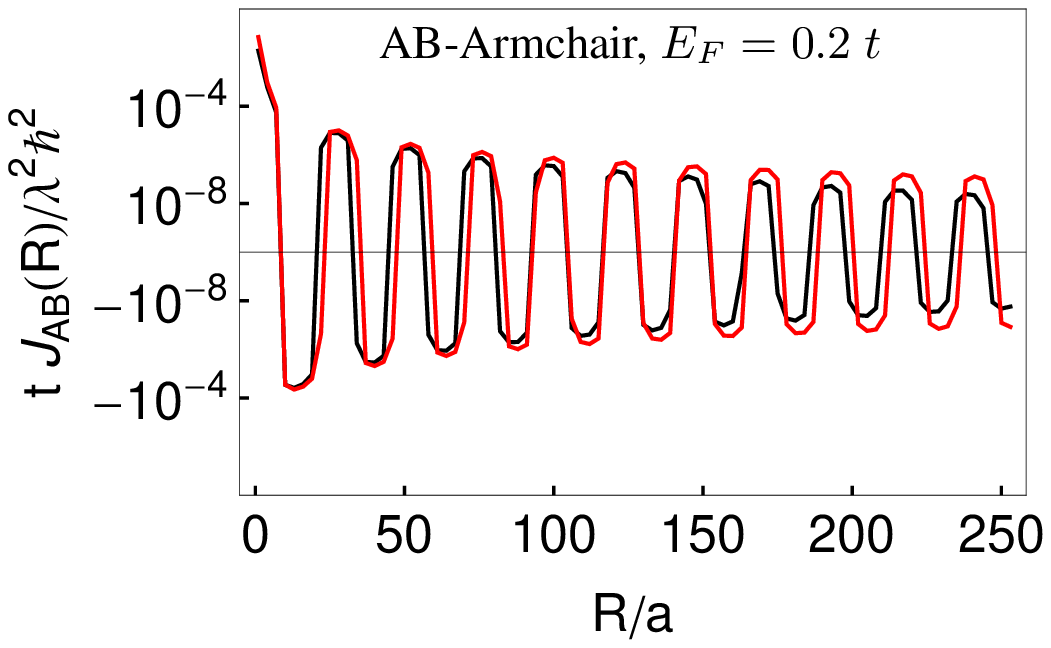}
\caption{(Color online) RKKY interaction for several cases. Black solid lines are the numerical results for  the full tight-binding band structure, while the red lines indicate the analytical results, Eqs. (\ref{JAAdoped}) and (\ref{JABdoped}), for the linear bands.
}
\label{Fig-J}
\end{figure}

Note that in the expression for the RKKY interaction Eq. (\ref{JAAdoped}) the Fermi momentum term in the square bracket depends only on the magnitude of the distance $R$, while the Dirac cone term $J^0_{AA}(R)$ depends on its direction as well, which makes the interaction direction dependent. Here $K$ and $K'$ are any two adjacent Dirac points in the Brillouin zone. It is easy to see that while the oscillatory factor $1 + \cos ((K-K^\prime) \cdot R)$ repeats in triplets as 2, 1/2, 1/2, ... with distance $R$ along the zigzag direction, it is always two for the armchair direction, so that $J_{AA}$ changes smoothly along the armchair direction but not for the zigzag direction [Fig. (\ref{Fig-J})].

%

\begin{figure}[tp]
\centering
\includegraphics[width=6.0cm]{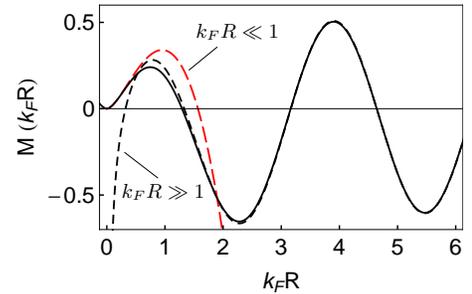}
\caption{(Color online) The Meijer G-function  as a function of $k_FR$ (black solid line) versus its asymptotic expansions (Eqs. \ref{MGS} and \ref{MGL}).
}
\label{MG}
\end{figure}

One is often interested in the long-distance behavior of the RKKY interaction and this may be obtained from the asymptotic behavior of the Meijer G-function $M(x)$. We find using standard tables\cite{Meijer} that
\begin{equation}
\lim _{x \rightarrow 0}M(x)=\frac{4 x ^2 [1-3 \gamma-3 \ln(x/2)]}{9 \sqrt{\pi}}
\label{MGS}
\end{equation}
\begin{equation}
\lim _{x \rightarrow \infty}M(x)=-\frac{\pi -2 \cos(2 x)-8 x  \sin(2 x)}{8 \sqrt{\pi} x},
\label{MGL}
\end{equation}
where $\gamma \approx 0.577$ is the Euler-Mascheroni constant. These functions are plotted in
Fig. (\ref{MG}). We note that with this asymptotic dependence, the
square bracket in  Eq. (\ref{JAAdoped}) becomes one for $k_F =0$, so that the RKKY interaction becomes the same as for the undoped case as it must.

Contrary to $J^0_{AA}(R)$ that always shows ferromagnetic coupling for the moments on the same sublattices owing to the particle-hole symmetry\cite{Saremi}, the oscillatory behavior of $M(k_FR)$ leads to the oscillations of $J_{AA}$ between ferromagnetic and anti-ferromagnetic interactions.
From Eqs. (\ref{MGL}) and (\ref{JAAdoped}), we obtain the long-distance behavior
\begin{equation}
\lim _{k_FR \rightarrow \infty} J_{AA}(R)=J^0_{AA}(R) \pi^{-1}  [2 \cos(2 x_F)+8 x_F \sin(2 x_F)],
\label{JAAlrg}
\end{equation}
where $x_F = k_F R$.
Note that the distance dependence is $R^{-2}$ if  $k_F $ is non-zero, i. e., the same  as for the ordinary 2D electron gas\cite{RKKY-2D}.  If $k_F = 0$, the RKKY interaction reverts to the undoped case as seen from Eqs. (\ref{JAAdoped}) and (\ref{MGS}), so that
the distance dependence is now  $R^{-3}$.

It is worth mentioning that the correct result for $k_FR\gg1$ can only be found by evaluating the Meijer G-function for large arguments and not just by replacing the Bessel functions in  Eq. (\ref{IAA-doped}) by their large-argument ($z \gg |\nu^2 - 1/4|$) limits, viz.,
$
J_\nu(z)\approx  2^{1/2} (\pi z)^{-1/2}      \cos \left( z- \nu\pi/2 - \pi/4 \right) $ and
$Y_\nu(z) \approx 2^{1/2} (\pi z)^{-1/2}      \sin \left( z- \nu\pi/2 - \pi/4 \right) .
$
The latter approach happens to lead to the same functional form as in Eq. (\ref{JAAlrg}) but with incorrect coefficients because of the error made in the small $k_FR$ contribution  to the integral in Eq. (\ref{IAA-doped}). However, there is no such problem if $k_FR \ll 1$ and the short-range results can be found either way to yield
%
%
%
\begin{equation}
J_{AA}(R)=J^0_{AA}(R)  \{1+\frac{32(k_FR)^3}{9 \pi}[1-3 \gamma-3 \ln(k_FR/2)]\}.
\label{JAAsml}
\end{equation}
%

{\it Moments on different sublattices} --
For moments located on two different sublattices,  we proceed as before to obtain the susceptibility \cite{SashiM}
\begin{equation}
\chi_{AB} (0,R) = I_{AB}  (R) \times \{1 + \cos [(K-K^\prime) \cdot R+\pi-2 \theta_R] \},
\label{chi-BA}
\end{equation}
where
$
I_{AB} (R) = \frac{4}{\pi} \int_{-\infty}^{E_F}  dE \  {\rm Im}  \ [g_{AB} (R, E)]^2,
\label{IBA}
$
and
$
g_{AB}(R,E) = - 2 \pi E v_F^{-2} \Omega_{BZ}^{-1} K_1(-i E R/v_F).
$
Expanding the modified Bessel function $K_1$, the integral becomes
\begin{equation}
I_{AB} (R) = \frac{8 \pi^3 R^{-3} }{\Omega_{BZ}^2 v_F} [I_0+
                  \int_0^{k_FR} dz \ z^2 J_1 (z) Y_1 (z)],
\end{equation}
where $I_0 = -\int_0^\infty dy \ y^2 J_1 (y) Y_1 (y) = 3/16$\cite{Saremi,SashiM} is the contribution from the undoped part
and the remaining integral can again be expressed in terms of the Meijer G-function
\begin{equation}
\int_0^{k_F R} dz \ z^2 J_1 (z) Y_1 (z)= -\frac{k_F R}{2\sqrt{\pi}} M'(k_FR).
\label{IBA-doped}
\end{equation}
This leads to the final result
\begin{equation}
J_{AB}(R)=J^0_{AB}(R) \ [1-\frac{8k_F R}{3\sqrt{\pi}}M'(k_FR)],
\label{JABdoped}
\end{equation}
where $M'(k_FR)=G_{2,4}^{\,2,1} \!\left( \left. \begin{matrix} \frac{1}{2},\frac{3}{2} \\ 1,2,0,\frac{-1}{2} \end{matrix} \; \right| \, k_F^2 R^2  \right)$  and the undoped exchange interaction is
$J^0_{AB}(R) = 3C \times (R/a)^{-3}  \{1 + \cos [(K-K^\prime) \cdot R+\pi-2\theta_R] \}.
$
%

%
\begin{figure} [tp]
\centering
\includegraphics[width=6.0cm]{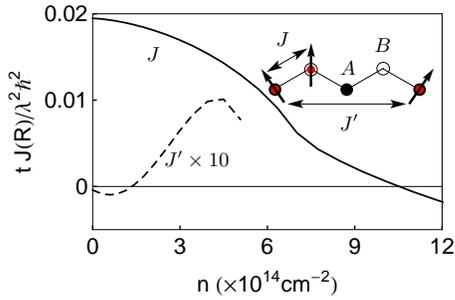}
\caption{(Color online) Switching the exchange interaction between ferro and antiferro by changing the carrier density in graphene with gate voltage. The larger the distance between the moments, the earlier is the switching,
which is controlled by $k_F R$.
}
\label{switching}
\end{figure}
%
\begin{figure}
\centering
\includegraphics[width=7.0cm]{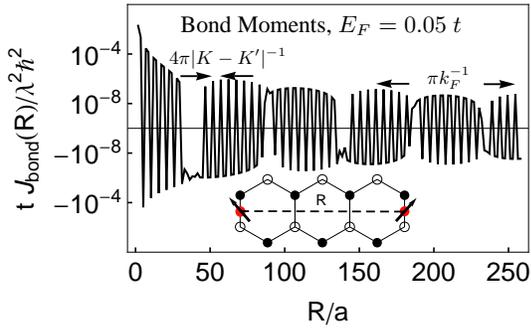}
\caption{(Color online) Beating pattern of the RKKY interaction for bond-centered moments separated along the zigzag direction.}
\label{Jbond}
\end{figure}

It has been demonstrated that the dopant carrier concentration in graphene can be controlled by a gate voltage or chemical doping.\cite{Novoselov} This raises the interesting possibility of switching the magnetic interaction between ferro and antiferro. This is illustrated in Fig. \ref{switching}, where the exchange interactions  were evaluated  using the full tight-binding bands as a function of $k_F$ and the carrier density is given by  $n=k_F^2/\pi$ in the linear-band region.

{\it Bond moments and the beating of $J$} -- For moments located on the bond center,
the interaction  is of the form
$
{\cal H}_{\text{int}} = - \lambda  \vec S \cdot \sum_{p} \vec s_{p},
$
where the summation is over the two adjacent atoms.
The exchange interaction becomes the sum of  the site interactions:
$
J_{\text{bond}} = 2J_{AA}+J_{BA}+J_{AB}
= 8 \pi^{-1} C (R/a)^{-3} \times  \{2 \cos(2 k_F R)-
3 \cos(2 k_F R) \cos [(K-K^\prime) \cdot R]-4 k_F R \  \sin(2 k_F R) \cos [(K-K^\prime) \cdot R] \}.
$
The resulting beating pattern of the RKKY interaction is shown in Fig. (\ref{Jbond}), which can be controlled by the gate voltage.

Finally, we note that Table I is valid both for electrons and holes because of the particle-hole symmetry.\cite{Saremi}
Mathematically, this follows from the fact that the net contribution to the susceptibility  from a symmetric range of energy is zero as may be seen by taking the integral in Eq. (\ref{chi}) from $-\varepsilon$ to $\varepsilon$ and by using the symmetry: $G^0_{\alpha \beta} (0, R, E) = G^0_{ \beta \alpha } (R, 0, E)$ and the fact that
the product $ {\rm Im} \ G^0_{\alpha \beta} (0, R, E) \times {\rm Re} \ G^0_{\alpha \beta} (0, R, E)$ is an odd function of energy.\cite{Horiguchi,Nanda-Vacancy}

In summary, we provided analytical results for the RKKY interaction in graphene in the linear-band approximation and showed that these results agree with the numerical results obtained for the tight-binding bands if the Fermi momentum is small. The presence of the two characteristic momenta, viz., the Dirac cone momentum $K-K'$ and the Fermi momentum $k_F$, leads to the unusual oscillatory features in graphene.

This work was supported by the U. S. Department of Energy through Grant No.
DOE-FG02-00ER45818. We thank Jet Foncannon for helpful discussions on the Meijer G-functions.


\end{document}